\documentclass[12pt]{article}
\usepackage{graphicx}
\usepackage{graphics}
\usepackage{url}
\usepackage{cancel}
\usepackage{amsmath}

\newcommand\pubnumber{CERN-PH-TH-2014-207}
\newcommand\pubdate{\today}

\def\napoli{CERN Theory Division,\\ CH-1211 Geneva 23,\\ SWITZERLAND}
\def\support{\footnote{Work supported by a Marie Curie Intra-European Fellowship of the European Community's 7th Framework Programme under contract number (PIEF-GA-2012-326948).}}

\def\Title#1{\begin{center} {\Large #1 } \end{center}}
\def\Author#1{\begin{center}{ \sc #1} \end{center}}
\def\Address#1{\begin{center}{ \it #1} \end{center}}

\newcommand\pubblock{\rightline{\begin{tabular}{l} \pubnumber\\
         \pubdate  \end{tabular}}}
\newenvironment{Abstract}{\begin{quotation}  }{\end{quotation}}
\newenvironment{Presented}{\begin{quotation} \begin{center} 
             PRESENTED AT\end{center}\bigskip 
      \begin{center}\begin{large}}{\end{large}\end{center} \end{quotation}}

\def\beq{\begin{equation}}
\def\eeq#1{\label{#1}\end{equation}}
\def\eeqn{\end{equation}}

\def\beqa{\begin{eqnarray}}
\def\eeqa#1{\label{#1}\end{eqnarray}}
\def\eeqan{\end{eqnarray}}

\let\bar=\overbar

\def\Dslash{\not{\hbox{\kern-4pt $D$}}}
\def\dslash{\not{\hbox{\kern-2pt $\del$}}}

\def\msb{{\bar{\ssstyle M \kern -1pt S}}}

\begin{document}
\begin{titlepage}
\pubblock

\vfill
\Title{Tauonic $B$ decays and Higgs mediated flavour violation}
\vfill
\Author{Andreas Crivellin\support}
\Address{\napoli}
\vfill
\begin{Abstract}
In these proceedings we review the impact of the tauonic $B$ decays $B\to\tau\nu$, $B\to D\tau\nu$ and $B\to D^*\tau\nu$ on Higgs mediated flavour violation. For this purpose we study a 2HDM with generic flavour structure (of type III). We find that despite the stringent constraints from FCNC processes~\cite{Crivellin:2013wna} sizable effects in tauonic $B$ decays are possible if top-related flavour violation is large and that the tensions with the SM predictions can be resolved~\cite{Crivellin:2012ye}, even when taking into account the recent CMS bounds on $A^0\to\tau^+\tau^-$. 
\end{Abstract}
\vfill
\begin{Presented}
$8^{th}$ International Workshop on the CKM Unitarity Triangle (CKM 2014),\\ Vienna, Austria,\\ September 8-12, 2014
\end{Presented}
\vfill
\end{titlepage}
\def\thefootnote{\fnsymbol{footnote}}
\setcounter{footnote}{0}

\section{Introduction}

Tauonic $B$-meson decays are an interesting probe of new physics: they test lepton flavor universality satisfied in the Standard Model (SM) and are sensitive to charged currents coupling proportional to the mass of the involved particles (e.g. Higgs bosons) due to the heavy $\tau$ lepton involved. An analysis of the semileptonic $B$ decays $B\to D\tau\nu$ and $B\to D^*\tau\nu$ using the full available data set was carries out by BaBar  \cite{BaBar:2012xj}. For the ratios
\begin{equation}
{\cal R}(D^{(*)})\,=\,{\cal B}(B\to D^{(*)} \tau \nu)/{\cal B}(B\to D^{(*)} \ell \nu)\,,
\end{equation}
in which the hadronic uncertainties, related to form factors, drop out to a large extent, BaBar obtaines the following results:
\begin{eqnarray}
{\cal R}(D)\,=\,0.440\pm0.058\pm0.042  \,,\qquad {\cal R}(D^*)\,=\,0.332\pm0.024\pm0.018\,.
\end{eqnarray}
Here the first error is statistical and the second one is systematic. Comparing this to the SM predictions
\begin{eqnarray}
{\cal R}_{\rm SM}(D)\,=\,0.297\pm0.017 \,, \qquad {\cal R}_{\rm SM}(D^*) \,=\,0.252\pm0.003 \,,
\end{eqnarray}
we see that there is a discrepancy of 2.2\,$\sigma$ for $\cal{R}(D)$ and 2.7\,$\sigma$ for $\cal{R}(D^*)$ and combining them gives a $3.4\, \sigma$ deviation~\cite{BaBar:2012xj}. This might be considered as evidence for new physics in $B$-meson decays to tau leptons and is further supported by the measurement of $B\to \tau\nu$ from BaBar and BELLE 
\begin{equation}
{\cal B}[B\to \tau\nu]=(1.15\pm0.23)\times 10^{-4}\,,
\end{equation}
which is $1.6\, \sigma$ above the SM prediction when using $V_{ub}$ from a global fit of the CKM matrix \cite{Charles:2004jd}. 

The impact of physics beyond the SM on tauonic $B$ decays has been extensively studied both in the effective field theory \cite{genericNPBDtaunu} and in specific models of NP \cite{Kalinowski:1990ba}. In these proceedings we examine the effect of Higgs mediated flavour violation in taunoic $B$ decays. For this purpose we consider a 2HDM with generic flavour-structure, i.e. of type III and update the numerical results of Ref.~\cite{Crivellin:2012ye} by taking into account the new CMS bounds from $A^0\to\tau^+\tau^-$~\cite{CMS:2013hja}. 

\section{Taonic $B$ decays in 2HDMs}

\begin{figure}[ht]
\begin{center}
\includegraphics[width=0.49\textwidth]{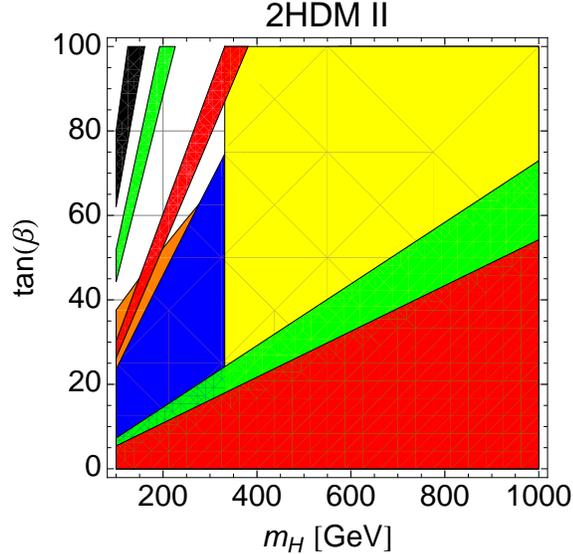}
\end{center}
\caption{Updated constraints on the 2HDM of type II parameter space from flavour observables \cite{Crivellin:2013wna}. The regions compatible with experiment are shown (the regions are superimposed on each other): $b\to s\gamma$ (yellow), $B\to D\tau\nu$ (green), $B\to \tau \nu$ (red), $B_{s}\to \mu^{+} \mu^{-}$ (orange), $K\to \mu \nu/\pi\to \mu \nu$ (blue) and $B\to D^*\tau \nu$ (black). Note that no region in parameter space is compatible with all processes. The tension originates from the destructive interference of the Higgs contribution with the SM one in $B\to D^*\tau \nu$ which would require very small Higgs masses and large values of $\tan\beta$ not compatible with the other observables. To obtain this plot, we added the theoretical uncertainty linearly on top of the $2 \, \sigma$ experimental error.}
\label{fig:2HDMII}
\end{figure}

The SM contains one Higgs doublet and has one physical Higgs particle. In a 2HDM one introduces a second Higgs doublet and obtains four additional physical Higgs particles (in the case of a CP conserving Higgs potential): a neutral CP-even Higgs $H^0$, a neutral CP-odd Higgs $A^0$ and two charged Higgses $H^{\pm}$. In the 2HDM of type II one Higgs doublet couples only to up-quarks while the other one couples only to down quarks and charged leptons (as in the MSSM at tree-level). Assuming a MSSM-like Higgs potential one has only two additional degrees of freedom, the heavy Higgs mass $m_H$ (which is approximately the same for $H^0$, $A^0$ and $H^{\pm}$) and $\tan\beta$, the ratio of the two vacuum expectation values. In Fig.~\ref{fig:2HDMII} a summery of the flavour constraints on the 2HDM of type II is shown. As one can see, it is not possible to explain $B\to D^*\tau\nu$ at the two $\sigma$ level without violating the other flavour bounds and the CMS constraint from $A^0\to\tau^+\tau^-$ \cite{CMS:2013hja}.

Therefore, we turn to a 2HDM with the most general Lagrangian for the Yukawa interactions (i.e. the 2HDM of type III). In the physical basis with diagonal quark mass matrices the Yukawa Lagrangian is given by
\begin{equation}
i \left( \Gamma_{q_f q_i }^{LR\, H} P_R + \Gamma_{q_f q_i }^{RL\, H} P_L \right) 
\end{equation}
with
\begin{eqnarray}
{\Gamma_{u_f u_i }^{LR\, H_k^0} } &=& x_u^k\left( \frac{m_{u_i }}{v_u}
\delta_{fi} - \epsilon_{fi}^{u}\cot\beta \right) + x_d^{k\star}
\epsilon_{fi}^{u}\,, \nonumber\\[0.1cm]
{\Gamma_{d_f d_i }^{LR\, H_k^0 } } &=& x_d^k \left( \frac{m_{d_i
}}{v_d} \delta_{fi} - \epsilon_{fi}^{d}\tan\beta \right) +
x_u^{k\star}\epsilon_{fi}^{ d} \,,\nonumber \\[0.1cm]
{\Gamma_{u_f d_i }^{LR\, H^\pm } } &=& \sum\limits_{j = 1}^3
{\sin\beta\, V_{fj} \left( \frac{m_{d_i }}{v_d} \delta_{ji}-
  \epsilon^{d}_{ji}\tan\beta \right)\, ,}
\nonumber\\[0.1cm]
{\Gamma_{d_f u_i }^{LR\,H^ \pm } } &=& \sum\limits_{j = 1}^3
{\cos\beta\, V_{jf}^{\star} \left( \frac{m_{u_i }}{v_u} \delta_{ji}-
  \epsilon^{u}_{ji}\tan\beta \right)\, }\, .
 \label{Higgs-vertices-decoupling}
\end{eqnarray}
Here, $H^0_k=(H^0,h^0,A^0)$ and the coefficients $x_q^{k}$ are given by
\begin{equation}
x_u^k \, = \, \frac{1}{\sqrt{2}}\left(-\sin\alpha,\,-\cos\alpha,\,i\cos\beta\right) \,,\qquad
x_d^k \, = \,\frac{1}{\sqrt{2}}\left(-\cos\alpha,\,\sin\alpha,\,i\sin\beta\right) \, .
\end{equation}
\smallskip
where $\epsilon^q_{ij}$ parametrizes the non-holomorphic corrections which couple up (down) quarks to the down (up) type Higgs doublet\footnote{Here the expression ``non-holomorphic" already implicitly refers to the MSSM where non-holomorphic couplings involving the complex conjugate of a Higgs field are forbidden.}. In the MSSM at tree-level $\epsilon^q_{ij}=0$ (as in the 2HDM of type II) and flavour changing neutral Higgs couplings are absent. However, at the loop-level, the non-holomorphic couplings $\epsilon^q_{ij}$ are generated~\cite{Hamzaoui:1998nu}\footnote{For a complete one-loop analysis including the resummation of all chirally enhanced effects see Ref.~\cite{Crivellin:2011jt} and for a 2-loop SQCD calculation see Ref.~\cite{Crivellin:2012zz}.}.

\begin{figure}[ht]
\centering
\includegraphics[width=0.4\textwidth]{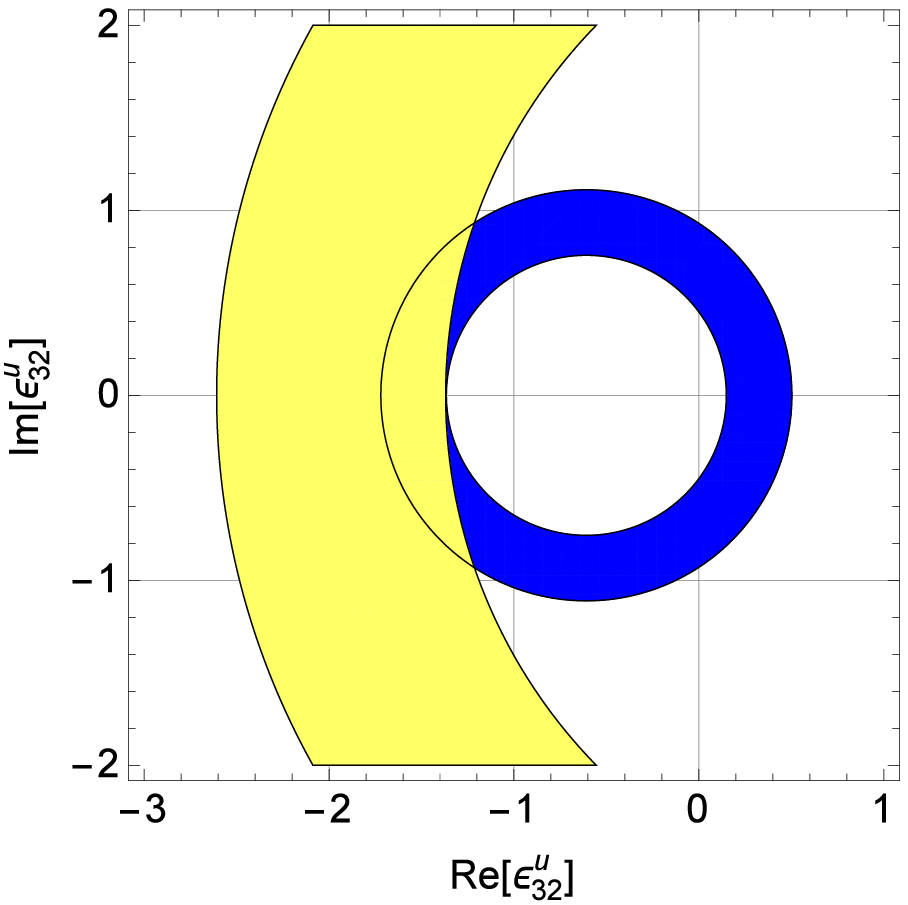}
\hspace{5mm}
\includegraphics[width=0.45\textwidth]{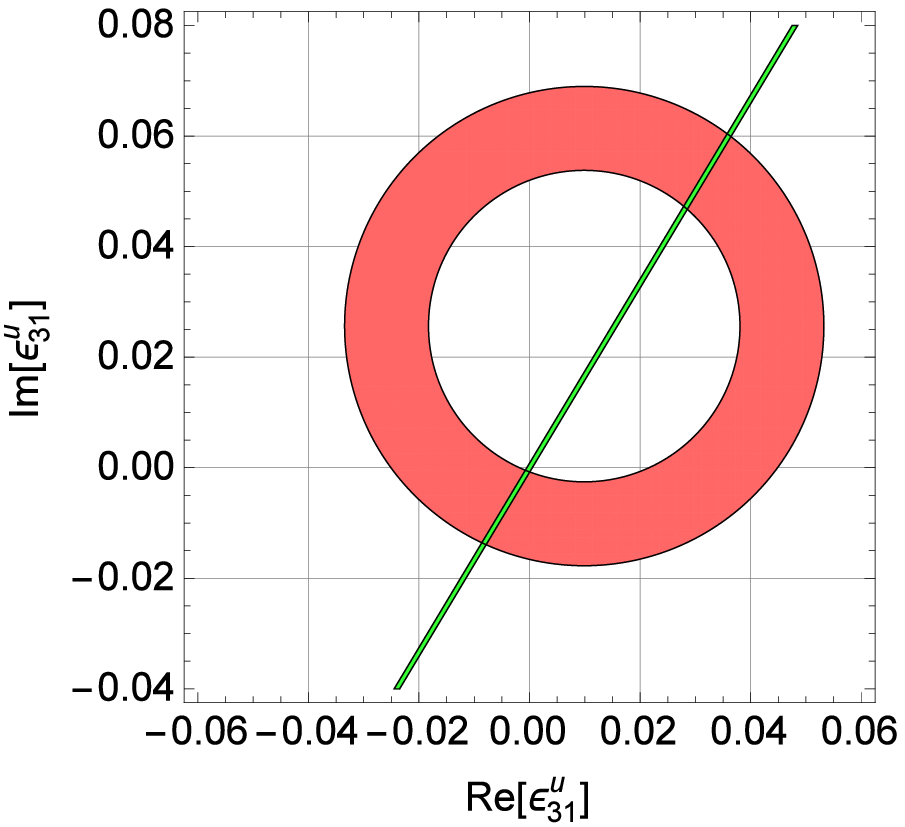}
\caption{Left: Allowed regions in the complex $\epsilon^u_{32}$-plane from $\cal{R}(D)$ (blue) and $\cal{R}(D^*)$ (yellow) for $\tan\beta=40$ and $m_H=800$~GeV. Right:  Allowed regions in the complex $\epsilon^u_{31}$-plane from $B\to \tau\nu$ (red) and the neutron EDM (green). \label{2HDMIII}}
\end{figure}

All $\epsilon^d_{ij}$ with $i\neq j$ are stringently constrained from $B_{q}\to\mu^+\mu^-$ and $K_L\to\mu^+\mu^-$ and $\epsilon^u_{12,21}$ is bounded by $D^0\to\mu^+\mu^-$. In addition $\epsilon^u_{13}$ ($\epsilon^u_{23}$) is stringently constrained from charged Higgs contributions to  $b\to d\gamma$~\cite{Crivellin:2011ba} ($b\to s\gamma$) and only $\epsilon^u_{31}$ ($\epsilon^u_{32}$) significantly effects $B\to \tau\nu$ ($\cal{R}(D)$ and $\cal{R}(D^*)$), even without any suppression by small CKM elements \cite{Crivellin:2013wna}. Furthermore, since flavor-changing $t\to u$ (or $t\to c$) transitions are not constrained with sufficient accuracy, we can only constrain these elements from charged Higgs-induced FCNCs in the down sector. However, since in this case an up (charm) quark always propagates inside the loop, the contribution is suppressed by the small Yukawa couplings of the up quark (charm quark). Thus, the constraints from FCNC processes are weak, and $\epsilon^u_{32,31}$ can be large. In fact, using $\epsilon^u_{32,31}$ we can explain $\cal{R}(D^*)$ and $\cal{R}(D)$ simultaneously \cite{Crivellin:2012ye}. In Fig.~\ref{2HDMIII} we see the allowed region in the complex $\epsilon^u_{32}$-plane, which gives the desired values for $\cal{R}(D)$ and $\cal{R}(D^*)$ within the $1\, \sigma$ uncertainties for $\tan\beta=40$ and $M_H=800$~GeV as allowed by $A^0\to\tau^+\tau^-$ \cite{CMS:2013hja}. Similarly, one can accommodate the measured values for $B\to \tau\nu$ by using $\epsilon^u_{31}$.

\section{Conclusions}

It is challenging to explain deviations from the SM in tauoinc $B$ decays with a UV complete model which respects all bounds from FCNC processes and LHC searches. While a 2HDM of type II cannot achieve this without violating the bounds from other flavour observables and $A^0\to\tau^+\tau^-$, a 2HDM of type III with large flavour violation in the up-sector can accommodate the BaBar measurements. In this model a sizable decay rate for $A^0,H^0\to t c$ is predicted. 
\medskip

{\bf Acknowledgements:} I thank the organizers for the invitation and the possibility to present these results.


\begin{thebibliography}{99}

\bibitem{Crivellin:2013wna}
  A.~Crivellin, A.~Kokulu and C.~Greub,
  Phys.\ Rev.\ D {\bf 87} (2013) 9,  094031
  [arXiv:1303.5877 [hep-ph]].
	
\bibitem{Crivellin:2012ye}
  A.~Crivellin, C.~Greub and A.~Kokulu,
  Phys.\ Rev.\ D {\bf 86} (2012) 054014
  [arXiv:1206.2634 [hep-ph]].
	
\bibitem{BaBar:2012xj}
  J.~P.~Lees {\it et al.}  [BaBar Collaboration],
  Phys.\ Rev.\ Lett.\  {\bf 109} (2012) 101802
  [arXiv:1205.5442 [hep-ex]].
	
\bibitem{Charles:2004jd}
  J.~Charles {\it et al.}  [CKMfitter Group Collaboration],
	Eur.\ Phys.\ J.\ C {\bf 41} (2005) 1  [hep-ph/0406184].  
	
	\bibitem{genericNPBDtaunu}	
  J.~F.~Kamenik and F.~Mescia,
  Phys.\ Rev.\ D {\bf 78} (2008) 014003
  [arXiv:0802.3790 [hep-ph]].
\\
  U.~Nierste, S.~Trine and S.~Westhoff,
  Phys.\ Rev.\ D {\bf 78} (2008) 015006
  [arXiv:0801.4938 [hep-ph]].
\\
  S.~Fajfer, J.~F.~Kamenik and I.~Nisandzic,
  Phys.\ Rev.\ D {\bf 85} (2012) 094025
  [arXiv:1203.2654 [hep-ph]].
  Y.~Sakaki and H.~Tanaka,
  ``Constraints on the charged scalar effects using the forward-backward asymmetry on $B\to D^{(*)}\tau\nu_{\tau}$,''
  Phys.\ Rev.\ D {\bf 87} (2013) 5,  054002
  [arXiv:1205.4908 [hep-ph]].

\bibitem{Kalinowski:1990ba}
  J.~Kalinowski,
  Phys.\ Lett.\ B {\bf 245} (1990) 201.
  W.~S.~Hou,
  Phys.\ Rev.\ D {\bf 48} (1993) 2342.\\
  A.~G.~Akeroyd and S.~Recksiegel,
  J.\ Phys.\ G {\bf 29} (2003) 2311
  [hep-ph/0306037].\\
  M.~Tanaka,
  Z.\ Phys.\ C {\bf 67} (1995) 321
  [hep-ph/9411405].
	\\
  T.~Miki, T.~Miura and M.~Tanaka,
  hep-ph/0210051.\\
  S.~Fajfer, J.~F.~Kamenik, I.~Nisandzic and J.~Zupan,
  Phys.\ Rev.\ Lett.\  {\bf 109} (2012) 161801
  [arXiv:1206.1872 [hep-ph]].
	\\
  A.~Datta, M.~Duraisamy and D.~Ghosh,
  Phys.\ Rev.\ D {\bf 86} (2012) 034027
  [arXiv:1206.3760 [hep-ph]].
  \\
  P.~Biancofiore, P.~Colangelo and F.~De Fazio,
  Phys.\ Rev.\ D {\bf 89} (2014) 095018
  [arXiv:1403.2944 [hep-ph]].
	\\
  X.~Q.~Li, Y.~D.~Yang and X.~B.~Yuan,
  Phys.\ Rev.\ D {\bf 89} (2014) 054024
  [arXiv:1311.2786 [hep-ph]].
	\\
  G.~Faisel,
  Phys.\ Lett.\ B {\bf 731} (2014) 279
  [arXiv:1311.0740 [hep-ph]].
	\\
  M.~Atoui, V.~Morénas, D.~Bečirevic and F.~Sanfilippo,
  Eur.\ Phys.\ J.\ C {\bf 74} (2014) 2861
  [arXiv:1310.5238 [hep-lat]].
	\\
  Y.~Sakaki, M.~Tanaka, A.~Tayduganov and R.~Watanabe,
  Phys.\ Rev.\ D {\bf 88} (2013) 9,  094012
  [arXiv:1309.0301 [hep-ph]].
	\\
  A.~Celis,
  PoS EPS {\bf -HEP2013} (2013) 334
  [arXiv:1308.6779 [hep-ph]].

\bibitem{CMS:2013hja}
  CMS Collaboration [CMS Collaboration],
  CMS-PAS-HIG-13-021.
	
\bibitem{Hamzaoui:1998nu}
  C.~Hamzaoui, M.~Pospelov and M.~Toharia,
  Phys.\ Rev.\ D {\bf 59} (1999) 095005
  [hep-ph/9807350].
	
\bibitem{Crivellin:2011jt}
  A.~Crivellin, L.~Hofer and J.~Rosiek,
  JHEP {\bf 1107} (2011) 017
  [arXiv:1103.4272 [hep-ph]].
	
\bibitem{Crivellin:2012zz}
  A.~Crivellin and C.~Greub,
  Phys.\ Rev.\ D {\bf 87} (2013) 015013
  [arXiv:1210.7453 [hep-ph]].
	
\bibitem{Crivellin:2011ba}
  A.~Crivellin and L.~Mercolli,
	Phys.\ Rev.\ D {\bf 84} (2011) 114005  [arXiv:1106.5499 [hep-ph]].  


\end{thebibliography}
\end{document}